\documentclass[fleqn,usenatbib]{mnras}

\usepackage{newtxtext,newtxmath}
\usepackage[T1]{fontenc}
\DeclareRobustCommand{\VAN}[3]{#2}
\let\VANthebibliography\thebibliography
\def\thebibliography{\DeclareRobustCommand{\VAN}[3]{##3}\VANthebibliography}

\newcommand\target{ASKAP\,J1755}
\newcommand\calebulp{ASKAP\,J1935+2148}
\newcommand\gcrt{GCRT\,J1745$-$3009}
\newcommand\gpm{GPM\,J1839$-$10}
\newcommand\gleamx{GLEAM-X\,J1627} % GLEAM-X\,J162759.5$-$523504.3
\newcommand\ilt{ILT\,J1101$+$5521}
\newcommand\chimelpt{CHIME\,J0630$+$25}
\newcommand\gleamxtwo{GLEAM-X\,J0704$-$37}

\usepackage{amsmath}
\usepackage{gensymb}
\usepackage{graphicx}
\usepackage{hyperref}
\graphicspath{{./}{figures/}}

\author[D. Dobie et al.]{Dougal Dobie,$^{1,2,3}$
Andrew Zic,$^{4}$
Lucy S. Oswald,$^{5,6}$
Joshua Pritchard,$^{3}$
Marcus E. Lower,$^{4}$
Ziteng Wang,$^{7}$
\newauthor
Hao Qiu,$^{8}$
Natasha Hurley-Walker,$^{7}$
Yuanming Wang,$^{1,2}$
Emil Lenc,$^{4}$
David L. Kaplan,$^{9}$
\newauthor
Akash Anumarlapudi,$^{9}$
Katie Auchettl,$^{2,10,11}$
Matthew Bailes,$^{1,2}$
Andrew D. Cameron,$^{1,2}$
Jeffrey Cooke,$^{1,2}$
\newauthor
Adam Deller,$^{1,2}$
Laura N. Driessen,$^{3}$
James Freeburn,$^{1,2}$
Tara Murphy$^{3,2}$
Ryan M. Shannon,$^{1,2}$
Adam J. Stewart,$^{3}$
\\
$^{1}$Centre for Astrophysics and Supercomputing, Swinburne University of Technology, Hawthorn, VIC 3122, Australia\\
$^{2}$ARC Centre of Excellence for Gravitational Wave Discovery (OzGrav), Hawthorn, Victoria, Australia\\
$^{3}$Sydney Institute for Astronomy, School of Physics, University of Sydney, Sydney, NSW 2006, Australia\\
$^{4}$Australia Telescope National Facility, CSIRO, Space and Astronomy, PO Box 76, Epping, NSW 1710, Australia\\
$^{5}$Department of Astrophysics, University of Oxford, Denys Wilkinson Building, Keble Road, Oxford OX1 3RH, UK\\
$^{6}$Magdalen College, University of Oxford, Oxford OX1 4AU, UK\\
$^{7}$International Centre for Radio Astronomy Research, Curtin University, Bentley, WA 6102, Australia\\
$^{8}$SKA Observatory, Jodrell Bank, Lower Withington, Macclesfield SK11 9FT, UK\\
$^{9}$Center for Gravitation, Cosmology, and Astrophysics, Department of Physics, University of Wisconsin-Milwaukee, P.O. Box 413, Milwaukee, WI 53201, USA\\
$^{10}$School of Physics, The University of Melbourne, Parkville, VIC 3010, Australia\\
$^{11}$Department of Astronomy and Astrophysics, University of California, Santa Cruz, CA 95064, USA\\
}

\title[Discovery of ASKAP\,J175534.9$-$252749.1]{A two-minute burst of highly polarised radio emission originating from low Galactic latitude}

\date{Accepted XXX. Received YYY; in original form ZZZ}

\pubyear{\the\year{}}

\begin{document}
\label{firstpage}
\pagerange{\pageref{firstpage}--\pageref{lastpage}}
\maketitle

\begin{abstract}
Several sources of repeating coherent bursts of radio emission with periods of many minutes have now been reported in the literature. These ``ultra-long period'' (ULP) sources have no clear multi-wavelength counterparts and challenge canonical pulsar emission models, leading to debate regarding their nature. In this work we report the discovery of a bright, highly-polarised burst of radio emission at low Galactic latitude as part of a wide-field survey for transient and variable radio sources. ASKAP\,J175534.9$-$252749.1 does not appear to repeat, with only a single intense two-minute $\sim$200-mJy burst detected from 60~hours of observations. The burst morphology and polarisation properties are comparable to those of classical pulsars but the duration is more than one hundred times longer, analogous to ULPs. Combined with the existing ULP population, this suggests that these sources have a strong Galactic latitude dependence and hints at an unexplored population of transient and variable radio sources in the thin disk of the Milky Way. The resemblance of this burst with both ULPs and pulsars calls for a unified coherent emission model for objects with spin periods from milliseconds to tens of minutes. However, whether or not these are all neutron stars or have the same underlying power source remains open for debate.
\end{abstract}


\section{Introduction} \label{sec:intro}
Searches for transient radio sources, independent of any multi-wavelength trigger, have traditionally been hindered by a combination of sensitivity and field-of-view. However this has changed with the advent of the Square Kilometre Array (SKA) Pathfinders -- the Murchison Widefield Array \citep[MWA;][]{2013PASA...30....7T,2018PASA...35...33W}, the Low Frequency Array \citep[LOFAR;][]{2013A&A...556A...2V}, MeerKAT \cite[][]{2016mks..confE...1J},  and the Australian SKA Pathfinder \citep[ASKAP;][]{2021PASA...38....9H}. All-sky surveys such as the Very Large Array Sky Survey \citep[VLASS;][]{2020PASP..132c5001L} and the Rapid ASKAP Continuum Survey \citep[RACS;][]{2020PASA...37...48M} have also led to the possibility of serendipitous transient searches. Millisecond timescale coherent transients such as fast radio bursts (FRBs) are now found regularly \citep[e.g.][]{2021ApJS..257...59C} and searches for synchrotron afterglows on timescales of weeks--years have also yielded several detections \citep[e.g.][]{2021Sci...373.1125D,2022MNRAS.510.3794D,2023ApJ...948..119D}.

Even with this proliferation of new discoveries, few radio surveys have probed intermediate timescales (seconds--hours) due to the computational and observational requirements of doing so. FRB searches can be conducted using standard pulsar backends and afterglow searches typically utilise a small number of deep images, but searching for radio variability on $\sim$minute timescales requires forming a much larger number of images which is computationally expensive. However, these searches are now becoming feasible -- \citet{2023MNRAS.519.4684D} reported the discovery of several radio sources with minute-timescale variability in a day-cadence transient survey, \citet{2024MNRAS.528.6985F} describe a search for radio transients on timescales of 8\,s, 128\,s and 1\,h, and \citet{2023MNRAS.523.5661W} reported the discovery of 38 radio variables on 15 minute timescales, including six rapid scintillators associated with a large plasma filament \citep{2021MNRAS.502.3294W}.

In addition, several sources of repeating coherent bursts with periods of minutes, much longer than canonical pulsar rotation periods, have now been reported. \chimelpt{} has a period of 7 minutes \citep{2024arXiv240707480D}, \gleamx{} has a period of 18.18 minutes \citep{2022Natur.601..526H}, \gpm{} has a period of 22 minutes \citep{2023Natur.619..487H} \calebulp{} has a period of 54 minutes \citep{caleb24}, \ilt{} has a period of $\sim 2.1$ hours \citep{2024arXiv240811536D} and \gleamxtwo{} has a period of $\sim$2.9 hours \citep{2024arXiv240815757H}. These six new discoveries are remarkably similar to the Galactic-centre radio transient \gcrt{}, which has a period of 77 minutes \citep{2005Natur.434...50H}. None of the sources have detectable persistent radio emission. Two (\ilt{} and \gleamxtwo{})have optical counterparts, while the others have no multi-wavelength counterparts \citep[][and the aforementioned discovery papers]{2008ApJ...687..262K,2022ApJ...940...72R}. For the purposes of this work, we consider these sources to be four members of the emerging class of ultra-long period sources (ULPs)\footnote{\citet{2022NatAs...6..828C} reported the discovery of a pulsar with a 76\,s period, although this has been suggested to have a canonical pulsar origin \citep[e.g.][]{2024ApJ...961..214R}.}. The physical origin of these sources is unclear, but the leading prospects are white dwarfs or neutron stars \citep[e.g.][]{2023ApJ...943....3T}. A number of scenarios have been proposed to explain the unusually long periods in these scenarios \citep[e.g., wind-dominated spin-down of a regular pulsar;][]{2024AN....34530176S}, while more exotic origins have also been proposed \citep[e.g.][]{2024PhRvD.109f3004B}.

In this work we present the discovery of a new radio transient source at low Galactic latitude. ASKAP\,J175534.87$-$252749.1 (hereafter \target{}) was detected as a single coherent burst lasting approximately 2 minutes, with high circular and linear polarisation. In Section \ref{sec:obs-and-results} we present the discovery and follow-up observations. In Section \ref{sec:analysis} we analyse the burst properties and in Section \ref{ref:discussion} we discuss possible origins and the implications of this discovery.

\section{Observations and Results}
\label{sec:obs-and-results}
\subsection{The Variables And Slow Transients (VAST) survey}
\label{subsec:vast}
The Australian Square Kilometre Array Pathfinder \citep[ASKAP;][]{2021PASA...38....9H} is a radio interferometer consisting of thirty-six 12\,m dishes located at Inyarrimanha Ilgari Bundara, the CSIRO's Murchison Radio-astronomy Observatory in Western Australia. Each dish is equipped with a phased array feed that is used to form 36 dual linear polarisation beams on the sky, giving ASKAP a field of view of approximately 30 square degrees at typical observing frequencies. All cross-correlations are recorded (providing full-Stokes information) in 10\,s integrations, and in the observations considered here, across 1\,MHz wide channels. All ASKAP imaging data are calibrated and imaged with {\sc ASKAPsoft}, which produces per-beam calibrated visibilities and full-Stokes images for the full field of view. Total intensity catalogues and noise maps are generated with {\sc selavy} \citep{2012PASA...29..371W}, with all data automatically uploaded to the CSIRO ASKAP Science Data Archive (CASDA\footnote{\url{https://research.csiro.au/casda/}}) and made public after passing scientific validation.

The Variables And Slow Transients (VAST) project \citep{2021PASA...38...54M} is a Survey Science Project that ASKAP will carry out across its first five years of operation. VAST observes a series of 42 fields (40 targeting the Galactic plane and one targeting each of the Magellanic clouds) on an approximately fortnightly cadence, with another 265 extragalactic fields observed approximately every two months. VAST observations use the {\textsc square\_6x6} footprint with 288\,MHz of bandwidth centered on 887.5\,MHz -- the same  as the Rapid ASKAP Continuum Survey \citep{2020PASA...37...48M}. The typical sensitivity across the full survey footprint is $250\,\mu$Jy, although it can reach $\sim 500\,\mu$Jy close to the Galactic plane due to bright sources, extended emission and source confusion.  

In this work we use the post-processed \citep{vpp_software} VAST data, which has been corrected for flux scale and astrometry relative to RACS and cropped to a central $6.3\degr\times6.3\degr$ square. We regularly ingest the post-processed data covering the period from into the VAST pipeline \citep{2022ASPC..532..333P} as it becomes available. The VAST pipeline takes images, noise maps and source catalogues from ASKAP observations, runs source association, carries out forced photometry where appropriate, and calculates source statistics. The resulting data can then be queried programmatically using {\sc vast-tools}, a python package designed for interacting with VAST data \citep{vast_tools}.

\subsection{Discovery}
\label{subsec:discovery}
As part of VAST pipeline testing we ran a search for sources with a single detection on an early subset of the VAST Galactic survey including observations through to early 2023. We selected sources that satisfied the following criteria:

\begin{itemize}
\item nearest neighbour distance larger than 1 arcminute;
\item average compactness (peak flux density divided by integrated flux density for each measurement) between 0.8 and 1.4;
\item number of relations equal to zero (i.e. the source is isolated).
\end{itemize}

These cuts remove most variable false positives such as extended multi-component sources and imaging artefacts close to bright sources. We also only consider sources with a maximum peak flux density $\geq 5$\,mJy and a signal-to-noise $\geq 10$ for observations in which they are detected. Using both parameters may seem unnecessary as they are comparable quantities, but doing so removes imaging artefacts in noisy areas (which would pass a peak flux density cut, but not a signal-to-noise cut) and noise spikes in areas where the source finder underestimates the noise (which would pass a signal-to-noise cut, but not a peak flux density cut).

We then searched for sources with a single detection and then used {\sc vast-tools} to determine whether that detection was circularly polarised. \target{} was the only source that satisfied all criteria in the original search, although several other sources have been detected in more recent searches, which will be presented in future work.

\target{} was detected in SB47253 on 2023-01-21 with a flux density of $25.9\pm 0.8$\,mJy and a circular polarisation fraction of $\sim$25\%. After applying astrometric corrections (see Appendix \ref{appendix:astrometry} for details), we measured a position of 17:55:34.9(1) -25:27:49.1(1), with uncertainties in the final digit quoted in brackets. This position is covered by two overlapping adjacent fields, meaning that there are 85 observations of the position in our sample.

\begin{figure*}
    \centering
    \includegraphics{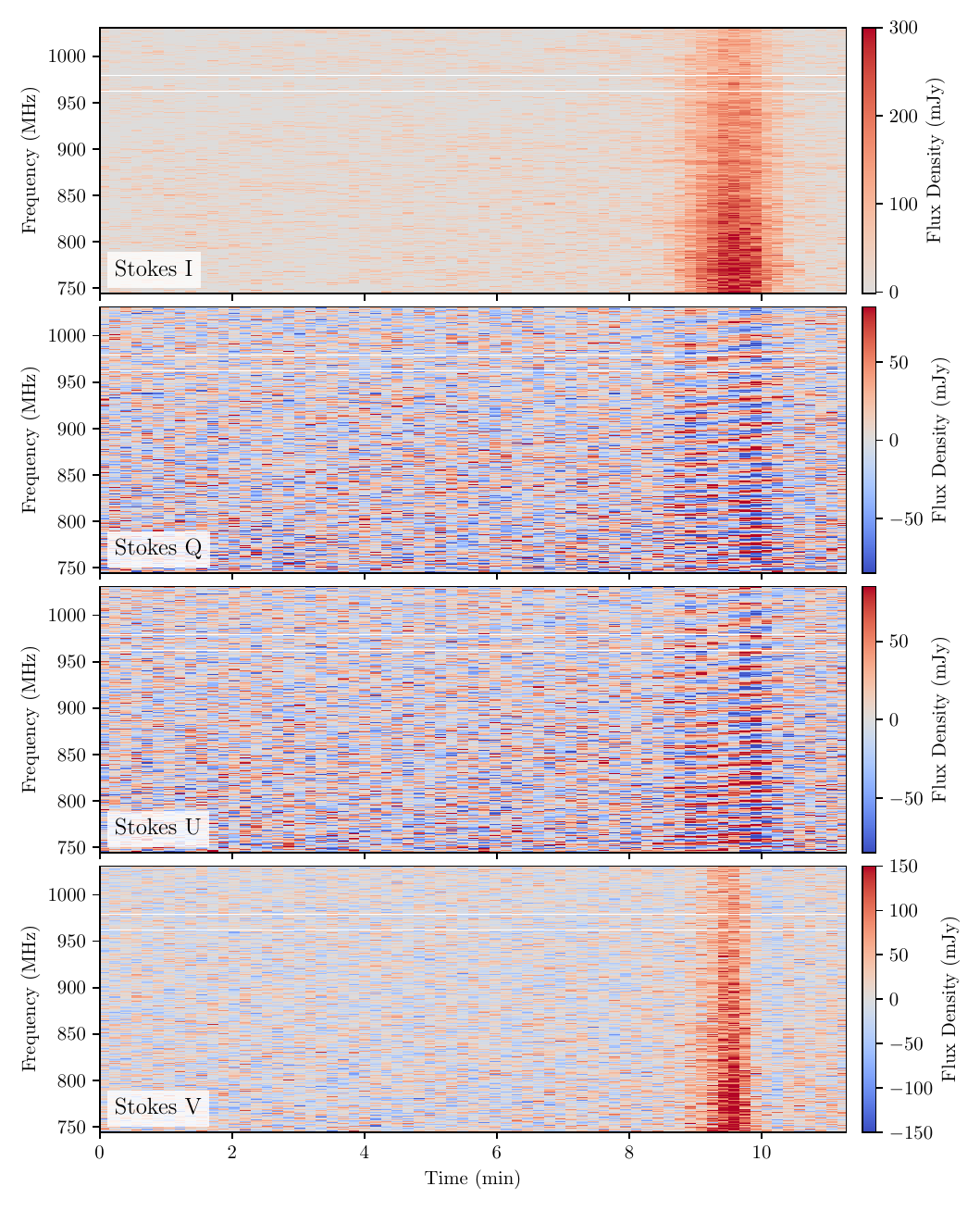}
    \caption{Dynamic spectra, showing flux density as a function of observing frequency (vertical axis) and time (horizontal axis) from the observation in which \target\ was discovered for all four Stokes parameters. Stokes Q and U data have not been corrected for Faraday rotation. The burst is detected in all polarisations at approximately 9.5 minutes into the observation.}
    \label{fig:full_ds}
\end{figure*}

\begin{figure}
    \centering
    \includegraphics{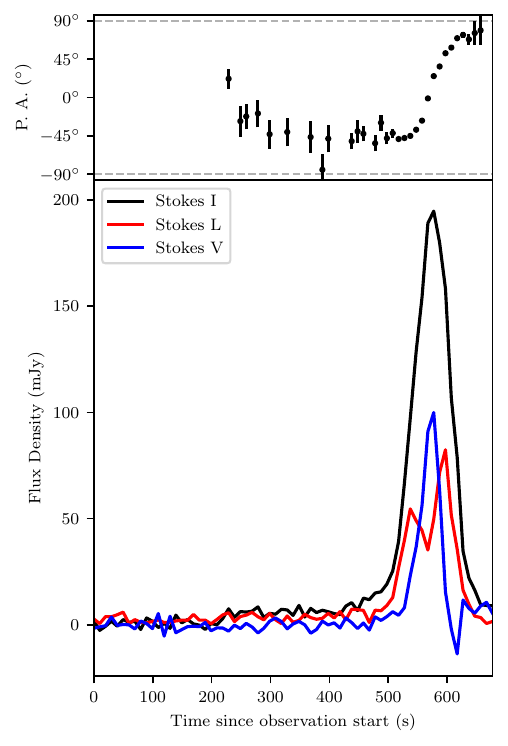}
    \caption{Bottom: frequency-averaged lightcurve for \target{}, in Stokes $I$ (black), de-biased linear polarisation corrected for Faraday rotation (``$L$''; red), and Stokes $V$ (blue). Top: the linear polarisation position angle, $\Psi$. Position angle measurements are only shown where they are statistically significant. Data are taken from SB47253, which started on 2023-01-21T00:51:45.}
    \label{fig:profile}
\end{figure}

We downloaded the calibrated measurement set of SB47253 beam 33 (the beam most sensitive at the position of the burst) from CASDA and generated dynamic spectra using {\sc DStools} \citep{joshua_pritchard_2024_13626183}. A detailed description of this process is available in Appendix \ref{appendix:dstools}. The resulting dynamic spectra (Figure \ref{fig:full_ds}) show that the observed radio emission consists of a short burst with both circular and linear polarisation components. We performed Faraday rotation measure (RM) synthesis \citep{2005A&A...441.1217B} and found an RM of $961\pm 45$\,rad\,m$^{-2}$. 

The dynamic spectrum shows no frequency-dependent time evolution and we are therefore able to average across time to create a spectral energy distribution (SED), and average across frequency to create a standard light curve timeseries. Averaging the dynamic spectrum for each Stokes parameter (after correcting the Stokes $Q$ and $U$ for Faraday rotation) results in the light curves shown in Figure \ref{fig:profile}. We also calculate the linear polarisation position angle, defined as 
\begin{equation}
    \Psi(t) = \frac{1}{2}\arctan\left(\frac{S_U(t)}{S_Q(t)}\right)
\end{equation}
where $\Psi$ is the polarisation angle in radians, and $S_Q(t)$ and $S_U(t)$ are the frequency averaged de-rotated flux density measurements of Stokes $Q$ and $U$ respectively. We measure a time above half-maximum of 80\,s for the Stokes $I$ (total intensity) light curve, although the emission is detectable above $5\sigma$ for 140\,s. Averaging the Stokes $I$ dynamic spectrum across the fourteen 10\,s time samples where the frequency-averaged lightcurve shows emission above $5\sigma$, we find the resulting SED is well fit by a power law with an extreme spectral index of $\alpha=-3.1\pm 0.1$, where $S_\nu \propto \nu^\alpha$, similar to the four known ULPs.

The light curve shows a weak leading flux density excess commencing $\sim$230\,s into the observation. This excess also shows substantial degrees of linear polarisation, with a position angle of $\sim -45^\circ$. Because the burst is located close to the end of the observation we are unable to determine whether this excess is symmetric in time or confined to before the burst. These scenarios correspond to a total burst duration of approximately 700 or 380 seconds respectively.

We also fitted the Stokes $I$ dynamic spectrum with a dispersed Gaussian model and find a tentative dispersion measure of $710^{+200}_{-180}$\,pc\,cm$^{-3}$ (see Appendix \ref{appendix:dm-fitting} for details), although we note that the inferred dispersion delay across our observing band is significantly less than the integration time. Using the YMW16 model for the Milky Way electron density \citep{2017ApJ...835...29Y}, this corresponds to a distance of $\sim 4.7$\,kpc and a luminosity of $\sim 5\times10^{21}$\,erg\,s$^{-1}$\,Hz$^{-1}$.  The  magnitudes of dispersion measure and rotation measure are consistent with Galactic plane lines of sight probed by e.g. radio pulsars (e.g. Figure S8 of \citealp{2023Sci...382..294R}).  

Based on the inferred burst distance, brightness, and duration we estimate a brightness temperature to be $T_b\approx 3 \times 10^{12}$\,K\,$D_{4.7}^2$, where $D=4.7 D_{4.7}$\,kpc is the distance to the burst source. This brightness temperature confirms the burst to have been produced by a coherent emission process. The brightness temperature, timescale, luminosity and spectral index are all comparable to the known ULP population \citep{2005Natur.434...50H,2022Natur.601..526H,2023Natur.619..487H,caleb24,2024arXiv240811536D,2024arXiv240815757H,2024arXiv240707480D}.

\subsection{Search for a multi-wavelength counterpart}
\label{subsec:multiwavelength}
\subsubsection{MeerKAT}
\label{subsec:meerkat}
We observed the position of the burst on 2023-05-28 with the L-band receiver on the MeerKAT radio telescope under a Director's Discretionary Time (DDT) proposal (proposal ID: DDT-20230525-DD-01). The observing block consisted of $11\times30$\,min scans on the target, interleaved with 60\,s scans of the phase calibrator J1833$-$2103. We used $2\times300$\,s scans of the radio galaxy J1939$-$6342 for bandpass and flux calibration, and a $200$\,s scan of J1331+3030 for polaristion calibration. We performed standard flagging, calibration and imaging with the IDIA MeerKAT pipeline\footnote{\url{https://github.com/idia-astro/pipelines}} to form a deep image, with a noise level of $\sim$25\,$\mu$Jy/beam in the region around the burst. The noise is dominated by the complex extended emission associated with the Galactic Centre. We detect no emission at the transient location, with a $3\sigma$ upper limit of $75\,\mu$Jy. We formed dynamic spectra using the same process as in Section \ref{appendix:dstools}. We find no evidence for any emission at the burst location in the 8\,s/1\,MHz dynamic spectrum with a $3\sigma$ upper limit of $750\,\mu$Jy. 

We queried the MeerKAT archive\footnote{\url{https://archive.sarao.ac.za/}} and found a series of observations with the UHF receiver covering the location of the burst under schedule block 20190215-0010. The schedule block consists of 9 separate pointings, of which two (N1R01C05 and N1R02C06) contain the burst location within the full-width-half-maximum of the primary beam. The corresponding time on source is 71\,minutes across an 8\,hour observation. The same bandpass, flux and gain calibrators were used, although no polarisation calibrator scan was conducted. We performed the same procedure as above to generate a dynamic spectrum and found no evidence for any emission at the burst location. The primary beam response at the burst location is $>0.7$, so we place a $3\sigma$ upper limit on repeat bursts within the dynamic spectrum of 1.1\,mJy.

\subsubsection{Murchison Widefield Array}
\label{subsec:mwa}
The burst position was also observed with the Murchison Widefield Array (MWA) at 185--215\,MHz as part of the Galactic Plane Monitoring program (GPM; project code G0080). The program is briefly described by \cite{2023Natur.619..487H} and will be fully described by Hurley-Walker et al. (in prep.)  There are 446 five-minute scans that cover the burst location spanning 2022-06-02 to 2022-09-08, corresponding to 74.3\,h of coverage in total. After subtracting a continuum model of the field, imaging was performed at a 4-s time cadence. No bursts are detected with a typical noise level of 50\,mJy\,beam$^{-1}$. This corresponds to a $3\sigma$ upper limit of $1.5\,$mJy at 888\,MHz, assuming a spectral index of $\alpha=-3.1$ based on the observed burst. 

\subsubsection{Murriyang}
\label{subsec:murriyang}

We performed follow-up observations with the Parkes/Murriyang radio telescope UWL receiver for 3.5\,h from 2023-04-13 20:30 UTC. 
Observations were conducted in search mode at centre frequency 1856 MHz with a bandwidth of 1024 MHz and 4096 channels (frequency resolution of 250kHz) and a time resolution of 64 $\mu$s. Observations were split into separate files, each containing 11.5 minutes of data.
We searched for periodic pulses using Heimdall and PRESTO and did not find detect any repetition burst activity from the observation.
The non-detection indicates an S/N>3 upper limit of 0.06 Jy with a repetition 95\% upper limit of $<0.85 \rm{hr^{-1}}$ \citep{gehrels1986_poisson}.

\subsubsection{Ultraviolet and X-rays}
\label{subsec:xray}
We searched for archival X-ray observations that overlap the position of \target{}. We find that the location of source was observed serendipitously by the \textit{XMM-Newton} Observatory \citep{2001A&A...365L...1J} $\sim256$ days after the source was discovered and by the Neil Gehrels Swift Observatory (\textit{Swift}; \citep{gehrels04}), $\sim$3792 days prior to discovery.  To place constraints on the presence on an X-ray source that may be associated with the radio transient, we reduced both the \textit{XMM-Newton} and \textit{Swift} X-ray Telescope data. 

\textit{Swift} observed the location of \target{} three times between MJD 56168 and MJD 56175 (ObsIDs: 00043737002,00043738001-00043738002). It simultaneously observed this location with the UltraViolet and Optical Telescope
\citep[UVOT;][]{roming05} and X-Ray Telescope \citep[XRT;][]{burrows05}. Due to the positioning of the UVOT and XRT, only 00043738001 and 00043738002 had overlapping UVOT observations of the source, while only the UVM2 filter was used for these observations. Aperture photometry was obtained using the {\sc UVOTSOURCE} package and a source and source-free background region of 5" and 20" respectively was used. We find no UV source at the location of the transient and obtain a 3 sigma AB magnitude upper limit of 21.5 in the UVM2 filter.

The source was observed using photon counting mode of the XRT and all observations were processed using the {\sc XRTPIPELINE} version 0.13.7, the most up-to-date calibrations and standard filters and screenings. As these observations were taken within a short period of time, we merged all three observations using \textsc{xselect} version 2.5b to place the strongest constaints on the X-ray emission at location of the transient prior to its discovery. Using a source region with a radius of 50" centered on the position of \target{} and a nearby source free background region, we find no significant X-ray emission from this location prior to the transient. Using this merged observation and assuming an absorbed powerlaw with a photon index of 2 and a Galactic column density of 1.37$\sim10^{22}$\,cm$^{-2}$ \citep{2016A&A...594A.116H}, we derive a 3$\sigma$ upperlimit for the absorbed (unabsorbed) flux of  1.4$\times 10^{-13}$\,erg\,cm$^{-2}$\,s$^{-1}$ (2.9$\times 10^{-13}$\,erg\,cm$^{-2}$\,s$^{-1}$) over the 0.3-10.0\,keV energy range. 

\textit{XMM-Newton} observed the location of \target{} once on MJD 60221 (ObsID: 0886090601). We reduced this observation using the \textit{XMM-Newton} Science System version 21.0.0, the most up to date calibration files and standard screening of events and flags as suggested by the \textit{XMM-Newton} Users Handbook. As \textit{XMM Newton} suffers from background flares and periods of high background, we filtered the data of these flares producing clean event files for our analysis. \textit{XMM-Newton} observations have three detectors corresponding to the MOS1,2 and PN. While the MOS detectors have a higher spatial resolution, due to the high sensitivity, and large effective area of the PN detector, we use the PN observation to constain the presence of X-ray emission coincident with the location of \target{}. We find no X-ray emission coincident with the location of the source after the detected radio flare. Using a circular region with a radius of 30" centered on the location of \target{} and a 120" source-free background region, we derive an 3$\sigma$ upperlimit to the count rate of 0.013 counts/sec in the 0.3-10.0\,keV energy band. Using the same absorbed powerlaw that we used for the \textit{Swift} data, we derive a 3$\sigma$ upperlimit to the absorbed (unabsorbed) flux of 6.6$\times 10^{-14}$\,erg\,cm$^{-2}$\,s$^{-1}$ (1.3$\times 10^{-13}$\,erg\,cm$^{-2}$\,s$^{-1}$) over the 0.3-10.0\,keV energy range. 

\begin{table}
    \centering
    \begin{tabular}{cccc}
        \hline\hline
        Survey & Band & AB Mag.\\
        \hline
        DECaPS & g & $>24.5$\\
        & r & $>24.0$\\
        & i & $>23.3$\\
        & z & $>22.5$\\
        & Y & $>21.4$\\
        \hline
        VVV & J & $>19.8$\\
        & H & $>19.0$\\
        & K & $>18.1$\\
        \hline\hline
    \end{tabular}
    \caption{Deepest optical and infra-red limits at the location of \target{} from archival DECam imaging and the VISTA Variables in the Via Lactea \citep[VVV;][]{2010NewA...15..433M}. We coadded all available frames at the burst location from the NOIRLab Astro Data Archive from 2013--2019 and VVV. These limits are not corrected for extinction, which may be as high as $\sim 75$ magnitudes in g-band.}
    \label{tab:optical_limits}
\end{table}

\subsubsection{Optical and infrared}
\label{subsec:optical}
The proximity of the burst to the Galactic plane means that the utility of optical and infrared observations is limited due to extinction from dust. Using the NED extinction calculator\footnote{\url{https://ned.ipac.caltech.edu/extinction_calculator}} we estimate there is $\sim 75$ magnitudes of extinction at $g$-band, decreasing to $\sim 7$ magnitudes at $K$-band \citep{2011ApJ...737..103S}, along this line of sight to extragalactic sources. The burst is likely Galactic, based on its Galactic latitude and tentative dispersion measure and hence the actual extinction to its position is lower than the above limits, but non-trivial to estimate accurately. Nevertheless, we have searched archival optical and infrared survey data for a potential counterpart to the burst.

We find no catalogued sources at the burst location, but do note the presence of a star $2.9\arcsec$ away from the burst position (RA=17:55:34.71, Dec.=-25:27:50.6; $i\sim 21.4$\,AB) in both the DECam Plane Survey \citep[DECAPS;][]{2023ApJS..264...28S} and PanStaRRS-1 surveys \citep[PS1;][]{2016arXiv161205560C}. Both surveys use stacked images that are not necessarily observed concurrently, with PS1 observations carried out from 2012--2013 and DECaPS observations carried out from 2016--2017. The position of the star is consistent between both surveys to within the $\sim$100\,mas uncertainties, and hence we infer a proper motion of $\lesssim 25\,$mas per year. We therefore rule out this star as the burst progenitor due to the $>10\sigma$ offset. For completeness we report the deepest optical and infrared limits available from archival data in Table \ref{tab:optical_limits}.

The burst location is covered by two major time-domain surveys - ZTF \citep{2019PASP..131a8002B,2019PASP..131g8001G} and PS1. ZTF observed the field 712 times between 2018-05-08 and 2023-06-30, to a typical sensitivity of 20.5 AB mag in the $g,r$ and $i$ filters, including minute-cadence monitoring consisting of 142 observations in the $r$ filter on 2019-06-02. PS1 observed the field 71 times between 2010-03-28 and 2014-06-27 in the $g,r,i,z$ and $y_{P1}$ filters to a typical sensitivity of 23 AB mag. Neither survey reports any detections at the burst location.

\section{Polarisation Analysis}
\label{sec:analysis}

\begin{figure}
    \centering
    \includegraphics[width=\linewidth]{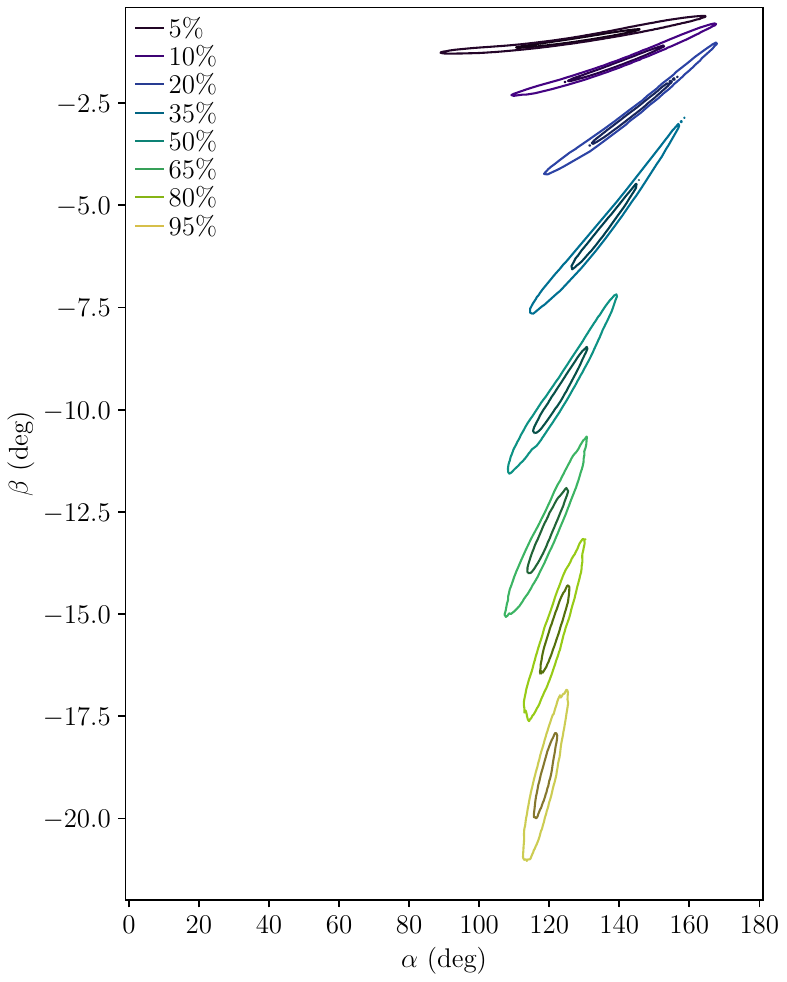}
    \caption{Resulting magnetic inclination angle ($\alpha$) and impact parameter ($\beta$) from RVM fits to ASKAP~J1755. Each set of contours shows the 68\% and 95\% confidence intervals, where the colour indicates the assumed pulse duty cycle.}
    \label{fig:rvm}
\end{figure}

\subsection{Rotating vector model}
\label{subsec:rvm}
The linear position angle swing across the profile shown in Figure~\ref{fig:profile} bears a striking resemblance to those seen among radio pulsars.
There, the S-shaped sweeps are interpreted under the rotating vector model (RVM) of \cite{1969ApL.....3..225R}, where the linear polarisation position angle (PA) corresponds to the sky-projected magnetic field direction of a rotating dipole on the plane of the sky.
Note that this formalism has also been previously applied in scenarios where the local magnetic topology of a possible higher-order multipolar field approximates a dipole (e.g., radio-loud magnetars; \citealt{2007MNRAS.377..107K, 2021MNRAS.502..127L}).
RVM fits to the PA swings of pulsars provide us with measurements of their magnetic and viewing geometries \citep{2023MNRAS.520.4801J}, where the PA as a function of rotation phase is modelled as
\begin{equation}\label{eqn:rvm}
    \tan(\Psi - \Psi_{0}) = \frac{\sin\alpha \sin(\phi - \phi_{0})}{\sin\zeta \cos\alpha - \cos\zeta \sin\alpha \cos(\phi - \phi_{0})}
\end{equation}
where $\Psi_{0}$ is the position angle at some reference pulse phase ($\phi_{0}$), $\alpha$ is the angle between the spin and magnetic axes of the star, and $\zeta$ is the inclination angle of the spin axis from our line of sight.
The angle of closest approach between the magnetic axis and our line of sight can be computed as $\beta = \zeta - \alpha$.

Typical RVM fits require a-priori knowledge of the pulsar rotation period.
If we assume that \target{} originated from a rotating object, such as a ULP, then we can infer its magnetic and viewing geometry via RVM fits to the position angle swing using various presumed pulse duty cycles. 
A similar approach was recently applied to an apparently non-repeating Fast Radio Burst that also displayed a curiously RVM-like position angle swing \citep{2024arXiv240209304M}.
In Figure~\ref{fig:rvm} we show the recovered posterior distributions of $\alpha$ and $\beta$ for eight different pulse duty cycles ranging between 5--95\% for a putative burst width of $\sim$380 seconds based on the time that any emission was detectable. Considering only the burst-like component would imply lower values of $\beta$, while assuming that the steady pre-burst component is symmetric around the burst would imply the opposite.
The recovered values of $\alpha$ are all broadly consistent regardless of the input duty cycle, with median values ranging between $119^{\circ}$ -- $142^{\circ}$ and substantial overlap at the 68\% confidence interval.
Conversely, the median magnitude of $\beta$ increases from $\sim 1^{\circ}$ to $\sim 19^{\circ}$ as the duty cycle is increased from 5\% to 95\%, as the corresponding decrease in position angle gradient across the burst requires a larger offset between the magnetic axis and our line of sight.
Very few radio pulsars have been found to possess $|\beta| > 15^{\circ}$, and those that do are subject to large measurement uncertainties \citep{2023MNRAS.520.4801J, 2023RAA....23j4002W}.
If \target{} arises from a population with $\beta$ similar to that of pulsars, then this model disfavours pulse duty cycles $>$80\%. The four published ULP sources have duty cycles ranging from 5--15\%, suggesting that if \target{} is a ULP then it has $|\beta| \lesssim 2.5^{\circ}$ based on Figure~\ref{fig:rvm}.

\subsection{Partially coherent mode addition}
\label{subsec:coherent_modes}
\begin{figure}
    \includegraphics[width=\columnwidth]{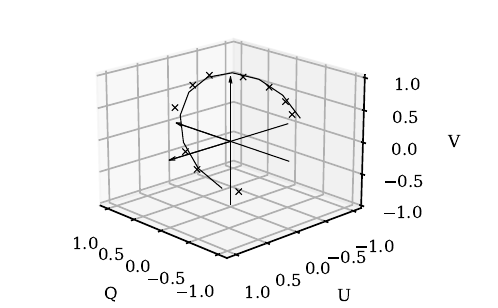}
    \caption{Black crosses: Poincar\'e sphere representation of the Stokes parameters $Q$, $U$ and $V$ as a fraction of total polarisation $P$ for the main burst of \target{}, from 527 to 617 seconds after the start of the observation. Black arc: output of modelling the burst as the partially coherent sum of two elliptical orthogonal modes with the phase offset between the modes increasing linearly with time. The arc shape replicates the behaviour of the data: full details of the model are given in the text.}
    \label{fig:poincare}
\end{figure}

Although the position angle swing appears to resemble an RVM curve, the rotating-vector-model interpretation of the burst polarisation does not consider the substantial circular polarisation present. \cite{Dyks2020} discussed how, for radio pulsars, a transition of polarisation state close to the Stokes $V$ pole on the Poincar\'e sphere can lead to distortion of the PA that appears to resemble an RVM swing. This motivates considering the polarisation state as a whole, including the circular polarisation: plotting the Stokes parameters $Q$, $U$ and $V$ as fractions of total polarisation $P$, we find that the polarisation state of the main burst follows an arc on the surface of the Poincar\'e sphere (see Fig. \ref{fig:poincare}). 

Similar arc shapes have been seen in the polarisation of radio pulsars and have often been interpreted as coherent mode transitions \citep{Edwards2004, Dyks2021a, Primak2022}. \cite{Oswald2023a} demonstrated that, for the radio pulsar population, the presence of strong circular polarisation is correlated with deviations of the position angle from an RVM curve. \cite{Oswald2023b} developed the partial coherence model to explain this as originating from two orthogonal modes combining together partially coherently, and partially incoherently, to generate simultaneously in the observed pulsar a rotation of the position angle, the presence of circular polarisation, and an overall depolarisation of the radio emission. They demonstrated that the partial coherence model can generate arcs on the Poincar\'e sphere if the phase offset between the two orthogonal modes is allowed to vary smoothly across the pulse. 

Motivated by the similarity between \target{} and these results for radio pulsars, we sought to use the partial coherence model of \cite{Oswald2023b} to account for the polarisation state of this burst. We find that the burst is best described by the partially-coherent addition of orthogonal modes that are also inherently elliptical. The relative amplitudes of these modes remain constant across the burst, as does the coherence fraction (the extent to which they combine coherently vs. incoherently). However, the phase offset between the two modes increases linearly with time to produce the varying polarisation state across the burst. Using the fully elliptical version of the partial coherence model \citep[see appendix A4 from][]{Oswald2023b} we are able to visually replicate the key polarisation features of the burst with the following constant parameters: $C = 0.58$, $\gamma = 0.8$, $\alpha = -1$, $\beta = 1$, $\chi = -35\degree$, $\psi = 35\degree$, and with the phase offset $\eta$ varying linearly with time between $-35\degree$ and $185\degree$. This generates the same arc shape on the Poincar\'e sphere. These parameters describe an inherent polarisation mode with a position angle of $-45\degree$ and an ellipticity angle of $17.5\degree$, which is then combined partially coherently with its orthogonal counterpart. In this model, the position angle is inherently flat, and equal to $-45\degree$, throughout the entire burst, not just for the previously described weak leading flux density excess. The rotation of polarization state we observe between 527 and 617 seconds comes entirely from combining the two modes with a varying phase offset. 

\begin{figure}
    \includegraphics{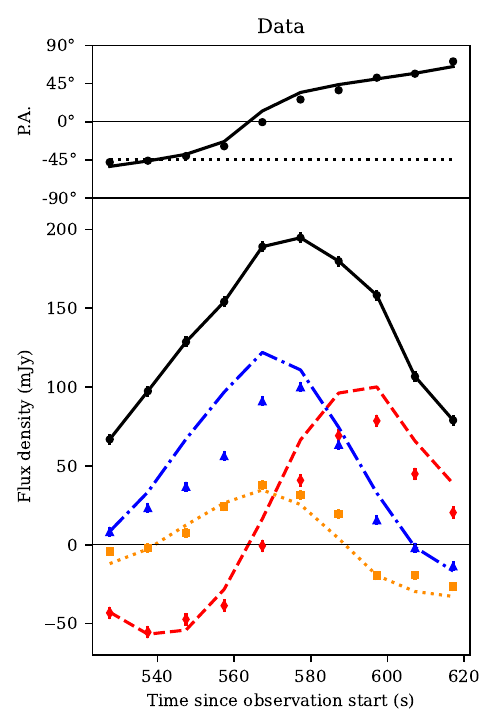}
    \caption{Polarised radio emission from \target{} around the time of the burst. The top panel shows the polarisation angle profile and the bottom panel shows the light curve for Stokes $I$ (black), $Q$ (orange), $U$ (red) and $V$ (blue). The markers in both panels show the observed data, while the lines show the model as described in Section \ref{subsec:coherent_modes}. Uncertainties on the observed data are included, but in most cases are smaller than the marker size. The model only contains information about relative mode intensity, so we scale the output by the total intensity of \target{} to generate Stokes parameters for comparison. In the PA plot, we mark the intrinsic model PA with a horizontal dashed line at 45$\degree$.}
    \label{fig:compareburst_datamodel}
\end{figure}

\begin{figure*}
    \includegraphics{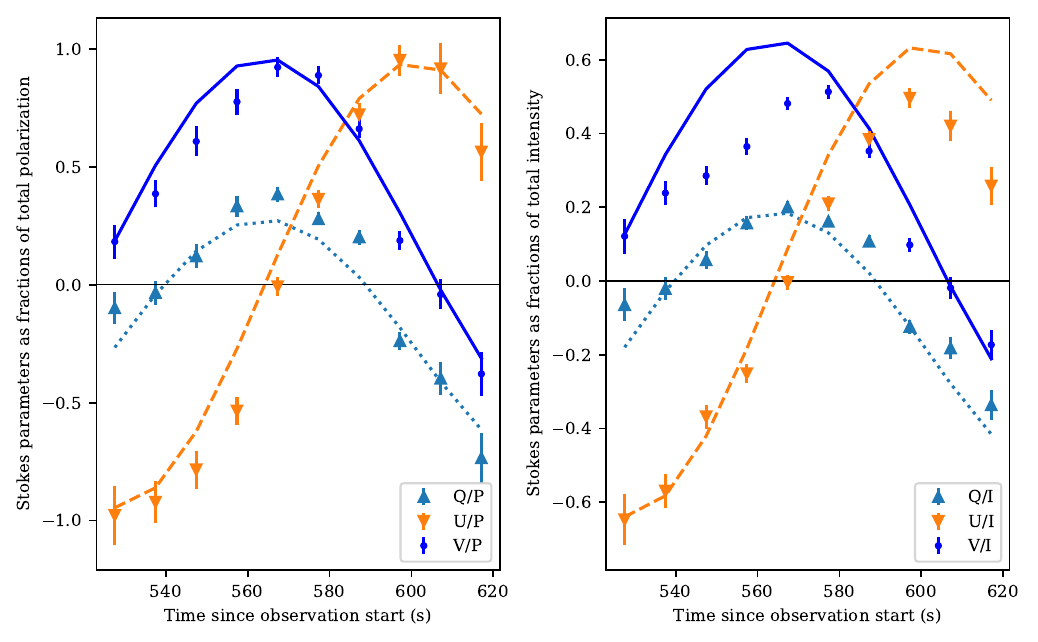}
    \caption{Left: Comparison of the Stokes parameters $Q$, $U$ and $V$ as a fraction of total polarisation $P$ of \target{} (shown with triangles and points) and the equivalent output parameters of the partial coherence model (shown with dotted, dashed and solid lines). Right: the equivalent comparison, but now comparing Stokes parameters as a fraction of total intensity $I$, rather than total polarisation $P$.}
    \label{fig:data_model_compare_Stokes}
\end{figure*}

Fig. \ref{fig:compareburst_datamodel} provides a visual comparison between \target{} and the output of the partial coherence model generated from these parameters. Comparing the Stokes parameters for the data and model more carefully in Fig. \ref{fig:data_model_compare_Stokes}, we see that the model does not completely replicate the Stokes parameters as a fraction of total intensity: the observed polarisation fraction is slightly lower than that modelled. This may mean that there is an additional source of unpolarized radio emission in addition to that generated by partially coherent mode addition, or alternatively it may be that the mathematical implementation of the partial coherence model is not completely accounting for all of the incoherent mode addition, possibly due to the model-simplifying choice of having only one coherence fraction parameter $C$ \citep[see appendix A2 of][]{Oswald2023b}. However, the model successfully recreates the Stokes parameters as a fraction of total polarisation: the simultaneous rotation of the PA and the presence of large amounts of circular polarisation are fully accounted for. Overall, the implication of being able to model the polarisation state of \target{} so successfully is that it must, at least in part, have been influenced by propagation of the radio waves through a birefringent medium, which would enable orthogonal modes to be produced, and to interact.

\section{Discussion}
\label{ref:discussion}
\subsection{Classification}
\label{subsec:classification}
The duration and luminosity of \target{} is broadly consistent with the current population of ULP sources, although the simple morphology of the total intensity burst differs with the sub-pulse structure observed in the current sample. We see no evidence for any repeat bursts in the other 80 VAST observations that cover this location from the start of the pilot survey in 2019 through to 2024-02-02, corresponding to 16 total hours on-source. We also find no repeat bursts in 6.7~hours of MeerKAT observations, corresponding to a combined duty cycle of $\sim 0.1\%$ at GHz-frequencies. Combining that with the lack of detections in 74.3 hours of the GPM survey (Section \ref{subsec:mwa}) we measure an overall duty cycle of $\sim 0.03\%$.

Such a low duty cycle may be intrinsic or may be the product of the bursts being intermittent or having a flat energy distribution. Given that \gpm{}, \gcrt{} and \calebulp{} are all highly intermittent, we interpret the low duty cycle as originating from intermittency. Assuming \target{} was active for 16 days either side of the detected burst (comparable to \gleamx{} being active for two one-month intervals, and motivated by the available VAST observations) we can conclusively rule out periods up to 43 minutes (implying a pulse duty cycle $\lesssim 5\%$), and most periods up to 2 hours (implying a pulse duty cycle $\lesssim 2\%$), based on the non-detections during that time. Motivated by \gcrt{} being active for at least 6 hours, we also consider a more conservative scenario, relying only on the observation in which it was detected and the observation of an adjacent overlapping field 27 minutes prior. In this scenario we can conclusively rule out periods up to 18.5 minutes and from 24--37 minutes (ruling out duty cycles $\gtrsim 10\%$ and from $\sim 8\%$ to $\sim 5\%$). Details of this analysis can be found in Appendix \ref{appendix-periodicity}. The range of potential duty cycles in either scenario are broadly consistent with the RVM parameters measured in Section \ref{subsec:rvm}. We conclude that the data are consistent with an intermittent ULP origin, although given the lack of repeat bursts we cannot conclusively determine this.

We also considered a stellar origin for \target{}. Highly-magnetized stars, such as ultra-cool dwarfs, M-dwarfs, RS CVn binaries, and magnetic chemically-peculiar stars produce coherent, highly-polarized bursts with timescales of milliseconds to hours (e.g. \citealp{2022ApJ...925..125D, 2019ApJ...871..214V, 2015Natur.523..568H, 2008ApJ...674.1078O, 2008PASA...25...94S}). Many of these stars produce periodic bursts at the stellar rotation period ($\sim$ hours), which is interpreted as the result of beamed emission originating from stable, field-aligned auroral current systems at the magnetic polar regions of the star \citep{2015Natur.523..568H, 2006A&ARv..13..229T, 2000ApJ...538..456E}. These bursts -- whether intermittent or periodic -- are usually highly circularly-polarized ($|V/I| > 50\%$; \citealp{2024MNRAS.tmp..161P, 2019ApJ...871..214V}), as expected for electron-cyclotron maser emission from a mildly-relativistic plasma. However, rare examples of elliptically polarised emission have recently reported  \citep{2017ApJ...836L..30L, 2019MNRAS.488..559Z, 2022ApJ...935...99B}, with fractional degrees of linear polarisation $\lesssim 40\%$. This contrasts with the polarimetric properties of \target{}, which instead exhibits a high degree of linear polarisation (peaking at $L/I > 60\%$), and a more moderate degree circular polarisation ($|V/I| \sim 40\%$) The typical spectral luminosities of stellar radio bursts are $10^{14}$--$10^{18}$\,erg\,s$^{-1}$\,Hz$^{-1}$ \citep{2024MNRAS.tmp..161P, 2024arXiv240407418D}, much lower than the estimated $\sim 5\times10^{21}$\,erg\,s$^{-1}$\,Hz$^{-1}$. 

Finally, the radio detections of stellar radio bursts can usually be associated with the responsible star in optical or infra-red imaging. Non-detection of an optical/infra-red counterpart therefore usually implies a distant star or cool star as the source of the emission. Taking the $J > 19.8$ infra-red limit (Table \ref{tab:optical_limits}), and the catalogue of nearby ultra-cool dwarfs from \citep{2008AJ....136.1290R}, we determine distance lower limits on low-mass stellar counterparts of $\sim 100$ -- $900$\,pc for L8 to M6 spectral types. Based on a radio-to-bolometric luminosity ratio $L_\text{rad}/L_\text{bol} \sim 10^{-5}$ \citep{2010ApJ...709..332B}, these distances would imply radio flux densities $< 90\,\mu\text{Jy}$ for a typical radio-emitting ultra-cool dwarf, which is a factor of $\sim 3000$ lower than the peak flux density of \target{}. We therefore rule out a low-mass stellar origin. 

No other known class of transient is consistent with the observed properties of the burst. For example, the timescale and presence of linear polarisation is inconsistent with a dwarf nova or cataclysmic variable interpretation \citep{2015MNRAS.451.3801C,2017MNRAS.467L..31M}. While the polarisation properties of the burst are comparable to canonical pulsars and magnetars, the duration is inconsistent. We cannot rule out the possibility that \target{} originates from a previously undiscovered class of transient radio source, but consider this an unnecessarily complex explanation. Instead our preferred interpretation for the burst is as a single pulse from an intermittent ULP source.

\subsection{A population confined to the Galactic Plane}
\label{subsec:rates}
\target{} is located at a Galactic latitude of $b=0.12\degr$ and the five of the seven ULP sources in the literature \citep{2005Natur.434...50H,2022Natur.601..526H,2023Natur.619..487H,caleb24} are all located within $\sim3$ degrees of the Galactic plane. The thin disk has a scale height of 279\,pc \citep{2023Galax..11...77V}, corresponding to an angular size of $\sim 2\degr$ at the Galactic Centre. Six of the seven ULP sources (all but \gcrt{}) have distance estimates, along with \target{} and hence their position within the Galaxy can be computed. \target{}, \gleamx{}, \gpm{}, \calebulp{} and \chimelpt{} all lie comfortably within the thin disk while \ilt{} and \gleamxtwo{} are located at heights of $|z|\approx 430$\,pc and $|z|\approx 330\,$pc above the Galactic plane respectively. While \gcrt{} does not have a distance constraint, its extremely low galactic latitude ($|b|\sim 0.5\degr$) suggests it also lies within the thin disk.

Interpretation of the positions of the existing sample of ULPs is subject to some level of selection bias -- e.g. \gcrt{} was discovered in a targeted search of the Galactic Centre and \gpm{} was discovered in a search of the Galactic plane covering $|b|<15\degr$. However, the VAST survey covers $\sim 1600\,\deg^2$ of the Galactic plane, with coverage extending up to $|b|\sim 10\degr$, along with 265 other extragalactic fields covering $\sim 10,000\,\deg^2$. Hence, our search is less affected by a potential positional selection bias than previous searches.

Approximately 85\% of the Milky Way's stars are located in the thin disk \citep{2020AJ....160...43A,2010IAUS..265..304A}, so a population of transients confined to the Galactic plane is not entirely unexpected -- indeed, it would almost be surprising if there was not one. However, the observed distribution is so far inconsistent with the distribution of classical pulsars which are found at a wider range of Galactic latitudes due to high-velocity kicks imparted during their formation \citep{2005MNRAS.360..974H}. It is also inconsistent with the observed distribution of white dwarfs, which spans a much wider range of galactic latitudes \citep{2021MNRAS.508.3877G}. In contrast, the observed magnetar population (which is presumably biased towards young, active objects) {\em is} found almost exclusively at very low Galactic latitudes, with a typical scale height of 20--31\,pc \citep{2014ApJS..212....6O}.

If the ULP population is confined to the thin disk, they must be associated with very young neutron stars, neutron stars with little or no kick, or a different class of progenitor. The first scenario is accessible if the stars rotation is rapidly decelerated shortly after birth, e.g. by fall-back accretion from the surrounding supernova remnant, thereby preserving their strong birth magnetic field \citep{2022ApJ...934..184R}. This might be necessary to generate the observed radio emission despite the slow spin periods. However, one might expect bright X-ray emission from such magnetar-like objects, and also that they would lie within visible supernova remnants, such as the compact object in RCW103 \citep{2016ApJ...828L..13R}. The large distances of the radio-selected ULPs may make X-ray emission difficult to detect, and confuse their surrounding supernova remnants with other Galactic emission too much to be detected, but this becomes increasingly untenable as further discoveries are made.

The second scenario is difficult to justify. For example, \cite{2023MNRAS.520.1872B} argue that the ULPs are instead older magnetars ($\sim10^{5.5}$--$10^{6}$\,years), with considerably different magnetic field and spin evolution to standard magnetars. Their modifications to these evolutionary scenarios have no predicted correlation with kick velocity. They therefore predict a population distribution similar to that of pulsars. The third scenario is an exciting prospect with as-yet unknown implications. Conclusively testing either the second or third scenarios requires a larger sample of ULP systems, both to confirm the galactic latitude dependence and to better understand the population properties in order to determine possible progenitors.

\subsection{Implications for pulsar and ULP emission models}
\label{subsec:implications}
The apparent consistency of the polarimetric properties of \target{} with either the RVM or the partial coherence model, as discussed in Section \ref{sec:analysis}, has important implications for models of pulsar and ULP emission mechanisms. Pulsar radio emission is thought to originate from a relativistic plasma within a rotating dipolar magnetosphere \citep[e.g.,][]{1975ApJ...196...51R}, with the resulting polarimetric phenomenology arising from a combination of viewing geometry and propagation effects within the magnetosphere \citep[e.g.][]{2023ApJ...952..151M, Oswald2023b, 1969ApL.....3..225R}. The shared polarisation morphology between \target{} and canonical pulsars therefore suggests that the emission from \target{} originates in similar circumstances. 

The burst width of \target{} is at least two orders of magnitude larger than that of canonical radio pulsars, and more comparable to that of the known ULP population. Under the ULP interpretation proposed in Section \ref{subsec:classification}, the pulsar-like polarisation properties of \target{} suggest that pulsar emission mechanisms may also be applicable to ULPs. However, at least one ULP  appears to lie below the commonly invoked ``death line'' \citep{2023Natur.619..487H}, where existing pulsar emission models predict that coherent radio emission should shut off \citep{1993ApJ...402..264C}. 
The existence of \target{} supports the proposition that the standard pulsar emission mechanism can operate up to spin periods of tens of minutes. If that mechanism powers both pulsars and ULPs then it requires substantial revision to explain the other members of the class. Alternatively, the ULP population may be made of several sub-classes with different origins and mechanisms that will be revealed as more are discovered, \`a la gamma-ray bursts. In this scenario, \target{} may arise from a canonical pulsar with a relatively high period derivative, while ULPs like \gpm{} that lie below the ``death line'' would produce radio emission via a different mechanism and have a non-neutron star origin.

\section{Conclusions}

The recent discoveries of multiple transient and highly variable radio sources with characteristic durations of minutes demonstrates that there is a gap in our understanding of the radio dynamic sky between the sub-second regime (e.g. pulsars, fast radio bursts) and the days-long regime (e.g. extragalactic synchrotron transients). The detection of a short burst of highly polarised radio emission, \target{}, presented in this work hints towards an unexplored population of minute-timescale variables combined to low galactic latitudes. While the burst does not appear to repeat, we argue that it is likely to be a member of the emerging class of "ultra-long period" sources. The similarities of \target{} with both ULPs and canonical pulsars suggests a relationship between the two classes and potentially a common emission mechanism.

While only a small number of ULPs have been discovered to-date, they are all tightly confined to the Galactic plane unlike canonical radio pulsars. The low galactic latitude of \target{} and the similarity of its polarisation properties with canonical pulsars should motivate a comprehensive search for minute-timescale variability around the Galactic Centre and at low Galactic latitudes more generally, in order to characterise ULPs, other transient or variable Galactic sources, and potentially find other new classes of object.

\section{Acknowledgments}
We thank M. Caleb, P. Uttarkar and C. Flynn for useful discussions.

Parts of this research were conducted by the Australian Research Council Centre of Excellence for Gravitational Wave Discovery (OzGrav), project numbers CE170100004 and CE230100016.

LSO and MEL are supported by the Royal Society International Exchange grant IES\textbackslash R1\textbackslash 231332. LSO also acknowledges the support of Magdalen College, Oxford. DK is supported by NSF grant AST-1816492.
NHW is supported by an Australian Research Council Future Fellowship (project number FT190100231) funded by the Australian Government. 

This scientific work uses data obtained from Inyarrimanha Ilgari Bundara / the Murchison Radio-astronomy Observatory. We acknowledge the Wajarri Yamaji People as the Traditional Owners and native title holders of the Observatory site. CSIRO’s ASKAP radio telescope is part of the Australia Telescope National Facility (\url{https://ror.org/05qajvd42}). Operation of ASKAP is funded by the Australian Government with support from the National Collaborative Research Infrastructure Strategy. ASKAP uses the resources of the Pawsey Supercomputing Research Centre. Establishment of ASKAP, Inyarrimanha Ilgari Bundara, the CSIRO Murchison Radio-astronomy Observatory and the Pawsey Supercomputing Research Centre are initiatives of the Australian Government, with support from the Government of Western Australia and the Science and Industry Endowment Fund.

This paper includes archived data obtained through the CSIRO ASKAP Science Data Archive, CASDA (\url{http://data.csiro.au})

The MeerKAT telescope is operated by the South African Radio Astronomy Observatory, which is a facility of the National Research Foundation, an agency of the Department of Science and Innovation.

\section*{Data Availability}
The observational data underlying this article is available on the respective telescope archives -- for example, calibrated ASKAP visibilities, images and source catalogues are available from CASDA (\url{http://data.csiro.au}) and MeerKAT visibilities are available from the SARAO web archive (\url{https://archive.sarao.ac.za/}). All other data will be shared on reasonable request to the corresponding author.

\bibliography{bibliography}{}
\bibliographystyle{mnras}

\appendix

\section{Generation of dynamic spectra}
\label{appendix:dstools}

We generated dynamic spectra and lightcurves using {\sc DStools} \citep{joshua_pritchard_2024_13626183}, a {\sc CASA} based python package to construct and post-process dynamic spectra from calibrated visibilities, and describe the details of each processing stage below.

We first applied some telescope-specific pre-processing. ASKAP visibilities processed with the ASKAPsoft pipeline are stored on a per-beam basis, but the MeasurementSet of each beam retains the centre of the 36-beam mosaicked field as the pointing centre rather than the true pointing centre of the individual beam. This causes a shift in the astrometric frame against which the phase centre is referenced if uncorrected, so we first updated the SB47253 beam 33 MeasurementSet to the correct pointing centre. ASKAPsoft also uses definitions of the Stokes parameters based upon the total flux of orthogonal correlations (e.g. $I = XX + YY$) while {\sc CASA} uses Stokes parameter definitions based on the average flux (e.g. $I = (XX + YY)/2$), so we multiplied all ASKAP visibilities by two in order to establish the correct flux scale in {\sc DStools} generated data products. 

We applied further automated flagging to the MeerKAT data using {\sc tfcrop} and {\sc rflag} with default settings. We then performed a custom flagging routine, removing all time-frequency chunks that had over half the baselines flagged, then flagging all baselines shorter than 100\,m and finally flagging all channels where over half the visibilities across the whole observation are flagged.

To form dynamic spectra we first built a model of all non-target sources in the field and subtracted this from the visibilities, such that the subtracted visibilities represent only the target emission and noise. We generated the field model from the ASKAP observation with {\sc tclean}, using the multi-term multi-frequency synthesis deconvolver with three Taylor terms to capture source spectral slope and curvature across the band. The MeerKAT measurement sets already include model components generated as part of the deep imaging described in \ref{subsec:meerkat}. For each measurement set we converted the model components to model visibilities using {\sc tclean} with the same parameters used during imaging and model construction, and subtracted them from the calibrated visibilities using the {\sc uvsub} task. We then rotated the phase-centre of the model-subtracted visibilities to the position of \target{} and averaged across all baselines, producing 2D time-frequency arrays of Stokes $XX$, $XY$, $YX$ and $YY$ fluxes. Using this procedure we extracted dynamic spectra from the ASKAP and MeerKAT observations at the native time and frequency resolution of 10\,s and 1\,MHz (ASKAP), 8\,s and 1\,MHz (archival MeerKAT) and 2\,s and 1\,MHz (MeerKAT DDT).

We then formed dynamic spectra for all Stokes parameters using the formalism
\begin{align}
    I &= \frac{1}{2}(XX+YY) \\
    Q &= \frac{1}{2}(XX-YY) \\
    U &= \frac{1}{2}(XY+YX) \\
    V &= \frac{i}{2}(YX-XY),
\end{align}
and constructed the total linear polarisation dynamic spectrum $L = \rm{Re}(Q) + i\rm{Re}(U)$ where $\rm{Re}(Q)$ and $\rm{Re}(U)$ are the real components of the Stokes $Q$ and $U$ dynamic spectra respectively, and $i$ is the imaginary unit. We also de-biased the linearly polarised intensity using the method described by \citet{2017A&A...600A..63M}. We performed Faraday rotation synthesis \citep{Brentjens2005} using {\sc RMclean} \citep{Heald2009, Heald2017} to extract a rotation measure of $\rm{RM} = 961\pm 45$\,rad\,m$^{-2}$ at the time of the burst peak, then corrected $L$ for Faraday rotation using
\begin{equation}
    L_{\rm intrinsic} = e^{-2i\lambda^2 \rm{RM}} L_{\rm observed}
\end{equation}
and extracted Faraday de-rotated $Q$ and $U$ dynamic spectra from the real and imaginary components of $L_{\rm intrinsic}$
\begin{align}
    Q_{\rm intrinsic} &= \rm{Re}(L_{\rm intrinsic}) \\
    U_{\rm intrinsic} &= \rm{Im}(L_{\rm intrinsic}).
\end{align}

\section{Inferred dispersion measure}
\label{appendix:dm-fitting}
\begin{figure*}
    \centering
    \includegraphics[width=\textwidth]{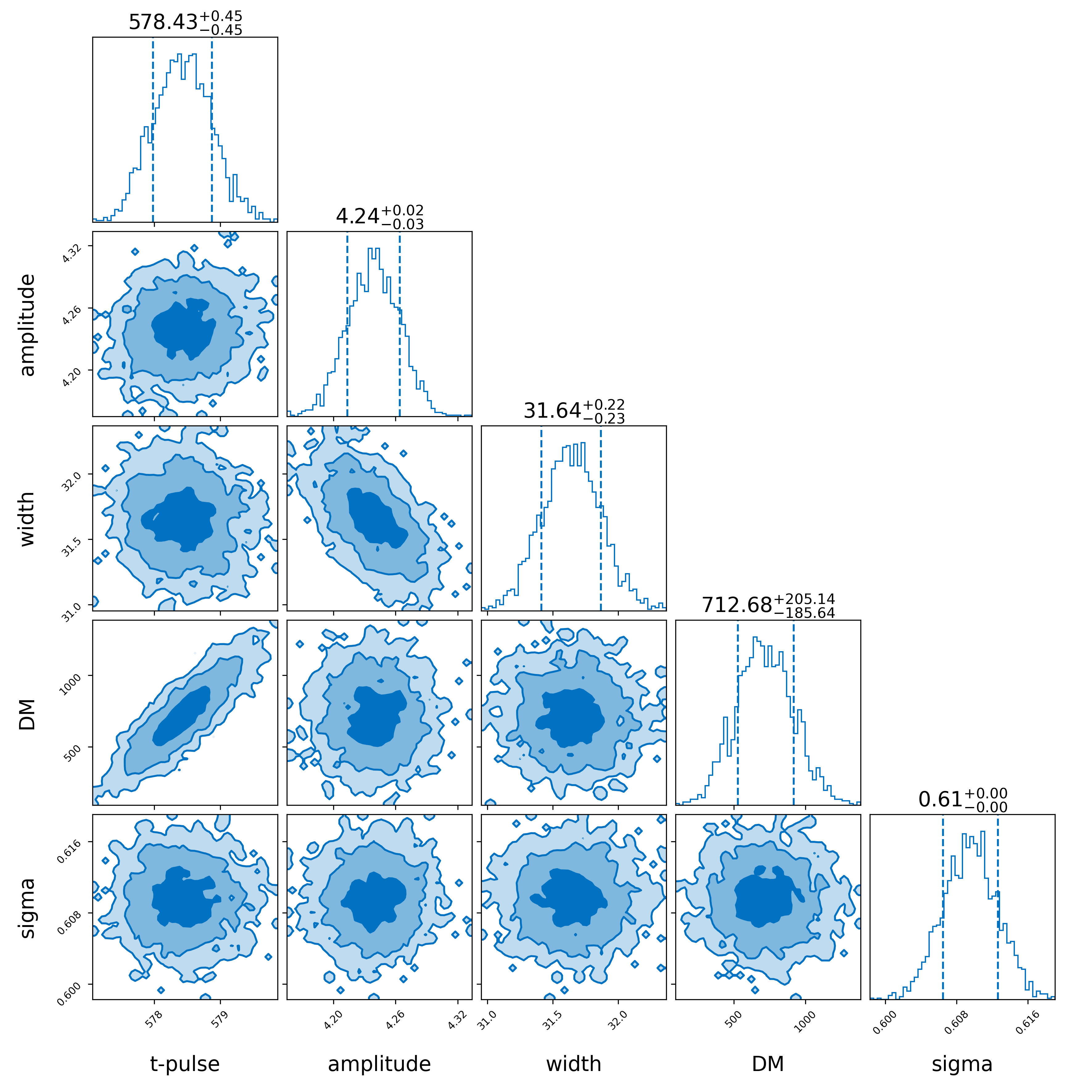}
    \caption{Corner plot for pulse profile fitting results, we display the pulse centre position(t-pulse), amplitude, pulse width and dispersion measure (DM). The r.m.s noise (sigma parameter) of the data set is also measured to check for correct for the amplitude units.}
    \label{fig:dm_fitting_corner_draft}
\end{figure*}

We fit the dynamic spectrum using a dispersed Gaussian pulse model based on the methodology in \citet{2020MNRAS.497.1382Q}, assuming no scatter broadening or smearing (which is expected to be on milli-second timescales and therefore not resolvable in our data). We assume uniform priors of 500-650 seconds since observation start for the central burst time $t_1$, 0-20 arbitary r.m.s units for the burst amplitude $a_1$, 10-100 seconds for the burst width $w_1$, and $0-3000$\,pc\,cm$^{-3}$ for the dispersion measure. We use 500 live points and apply nested sampling with Dynesty \citep{2020MNRAS.493.3132S} to fit for $\delta \rm{logz}<0.5$, which results in the posterior distributions shown in Figure \ref{fig:dm_fitting_corner_draft}. Specifically, we find a preferred DM of $\rm{7\pm2 \times 10^2\ pc/cm^3}$, corresponding to a time delay of approximately 2.6 seconds across the ASKAP band. Applying the model to the Stokes V data gives a similar result.

There are three potential issues with this fit.
\begin{enumerate}
    \item The inferred dispersion delay is significantly less than one time sample;
    \item The uncertainty in the dispersion measure is quite large; and
    \item The model does not account for any intrinsic frequency structure.
\end{enumerate}

\begin{figure*}
    \centering
    \includegraphics[width=\textwidth]{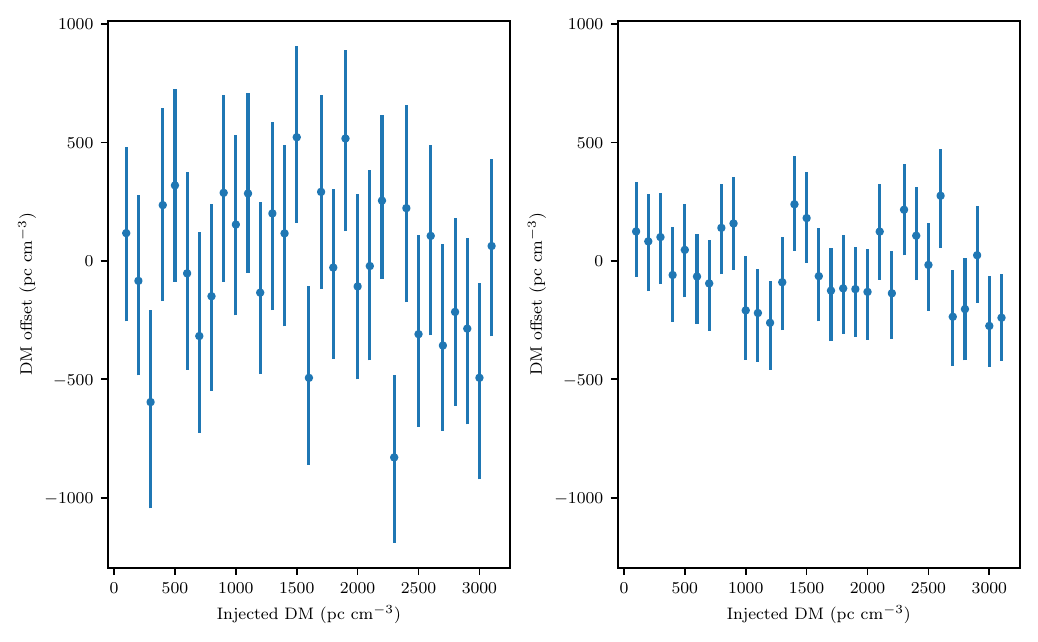}
    \caption{Dispersion measure (DM) recovery for simulated pulses as described in Appendix \ref{appendix:dm-fitting} The injected pulse DM 
    is shown on the x-axis, while the offset between the injected and measured DM values is shown on the y-axis. Left: DM recovery for a weak (S/N=5) pulse. Right: DM recovery for a strong (S/N=50) pulse.}
    \label{fig:dm_injection}
\end{figure*}

The first issue is not unexpected -- a 10\,s dispersion delay across the observed bandwidth corresponds to a dispersion measure of 2783\,pc\,cm$^{-3}$, which is in excess of the Milky Way contribution along this line of sight. To determine whether the recovered DM was sensible, we performed rigourous testing of this method by injecting Gaussian pulses into coarse-time-resolution white noise data with dispersion measures between 0-3000 $\rm{pc/cm^3}$. We considered two scenarios -- a weak (S/N=5) pulse and a strong (S/N=50) pulse. Figure \ref{fig:dm_injection} shows the results of our simulations, where we recover the injected DM in both scenarios.

The second issue stems directly from the first -- it is not possible to precisely infer the DM with such low time resolution. The third issue is difficult to address -- other ULPs show some complex frequency structure, but the origin of that structure is unclear and there are currently no models to reproduce it. Simultaneously, there is no obvious time or frequency structure in the dynamic spectrum light curve or spectral energy distribution.

All of these issues would be remedied by a detection of a repeat burst with a higher time resolution instrument, but as discussed in the main text, the source does not appear to repeat despite comprehensive observations. We therefore work with the available data and report a preferred DM of $\rm{7.1^{+2.0}_{-1.8}\times 10^2\ pc/cm^3}$, corresponding to a distance of $\sim4.7$\,kpc based on the YMW16 model \citep{2017ApJ...835...29Y}. For completeness we also report a 95\% confidence upper limit on the DM of 1000\,$\rm{pc/cm^3}$, which implies a distance of $\lesssim 6\,$kpc. The corresponding luminosity is $\sim 5\times 10^{21}$\,erg\,s$^{-1}$\,Hz$^{-1}$ in the former scenario and $\lesssim 8\times 10^{21}$\,erg\,s$^{-1}$\,Hz$^{-1}$ in the latter.

\section{Position measurement}
\label{appendix:astrometry}
ASKAP observations are not phase referenced, and hence by default, all VAST images have a systematic astrometric uncertainty of $\sim 1$ arcsecond \citep{2020PASA...37...48M}. Additionally, some VAST observations show systematic offsets of a few arcseconds in each beam - the origin of these offsets is not yet clear, but likely arises from a combination of the lack of phase referencing, ionospheric effects (exacerbated by high solar activity during the relevant observing period) and observing conditions. We correct for these offsets on a per-image basis as part of the VAST post-processing \citep{vpp_software}, but these corrections are not perfect as the offsets likely occur on a per-beam basis and the corrections we apply are relative to RACS, which is also impacted by the same issues.

In order to better correct for the expected astrometric offsets and improve the overall astrometric uncertainty we performed an astrometric correction relative to the phase-referenced MeerKAT observations on a per-beam basis. We used Aegean \citep{2012MNRAS.422.1812H,2018PASA...35...11H} to perform source-finding on the single-beam ASKAP image generated using the procedure in Section \ref{appendix:dstools} and the deep MeerKAT image of the field generated using the procedure in Section \ref{subsec:meerkat}. We filtered the list of sources from each image to exclude those with a peak flux density signal-to-noise ratio below 20, a compactness ratio (the ratio between the peak and integrated flux density) above 1.2 and those with a nearby source within $30\arcsec$ and $60\arcsec$ for MeerKAT and ASKAP respectively. This final filter removes source finding artefacts associated with bright sources, with the separation limits corresponding to approximately four beam-widths.

We then crossmatched both source lists with a $15\arcsec$ radius (approximately one ASKAP synthesised beam-width), resulting in 16 sources common to both images. For each common source we calculated the spherical offsets between its position in the ASKAP and MeerKAT images. We calculated the uncertainty in these offsets by adding the relevant positional uncertainties in quadrature. After combining the offsets of all common sources and weighting by the offset uncertainty, we measure a mean astrometric offset of $1.31\pm 0.08\arcsec$ in R.A. and $1.14\pm 0.07\arcsec$ in Declination.

For completeness we perform the same analysis using the NRAO VLA Sky Survey \citep[NVSS;][]{1998AJ....115.1693C} catalogue, which is an archival survey carried out at 1.4\,GHz. While it is less sensitive than both the MeerKAT and ASKAP images, it is a commonly used catalogue with well-understood astrometry. NVSS only provides integrated flux density values (not peak flux density), so we are unable to perform the same compactness ratio cuts as the MeerKAT-ASKAP comparison, and we use  a nearest-neighbour cut of $180\arcsec$ to reflect the larger synthesised beam. The analysis is otherwise identical to that outlined above. We find astrometric offsets of $2.14\pm 0.4\arcsec$ in R.A. and $1.4\pm0.4\arcsec$ in Declination based on 16 common sources, overall consistent to within $\sim 2\sigma$ with those obtained from the MeerKAT analysis. We adopt the MeerKAT values due to their higher precision.

After shifting the burst coordinates by the mean offset, we measure a final position of 17:55:34.9(1) -25:27:49.1(1), where the uncertainties are calculated by adding the statistical astrometric uncertainty from the ASKAP image and the uncertainty of the offset correction in quadrature.

\section{Periodicity constraints}
\label{appendix-periodicity}

\begin{figure*}
    \centering
    \includegraphics{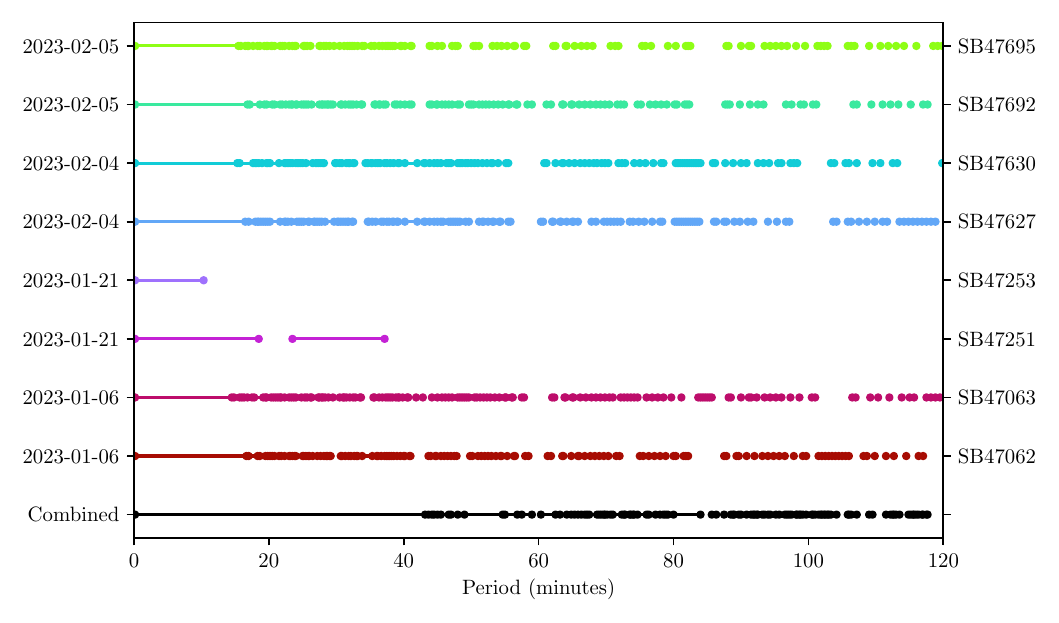}
    \caption{Constraints on the possible period of \target{} based on the detection time and non-detections in observations 16 days either side of the detection, assuming the source was active during that time. Each row corresponds to one observation (labelled with both observation date and Schedule Block) and each interval shows the period ranges that can be ruled out based on the non-detection of a repeat burst in that observation. The overall constraint from combining all eight observations under the assumption that the source was active for the entire interval is shown on the bottom row, showing that most periods up to 2 hours can be ruled out.}
    \label{fig:period-constraints}
\end{figure*}

\begin{table}
    \centering
    \begin{tabular}{rll}

    \hline\hline
     SBID & Observation start & Observation end \\
    \hline
    47062 & 2023-01-06 02:12:23 & 2023-01-06 02:24:23 \\
    47063 & 2023-01-06 02:26:11 & 2023-01-06 02:38:11 \\
    47251 & 2023-01-21 00:24:53 & 2023-01-21 00:36:53 \\
    47253 & 2023-01-21 00:51:45 & 2023-01-21 01:03:45 \\
    47627 & 2023-02-04 01:43:39 & 2023-02-04 01:55:39 \\
    47630 & 2023-02-04 02:26:07 & 2023-02-04 02:38:07 \\
    47692 & 2023-02-05 01:43:17 & 2023-02-05 01:55:17 \\
    47695 & 2023-02-05 02:26:35 & 2023-02-05 02:38:35 \\
    \hline\hline
    \end{tabular}
    \caption{Start and end times for all observations covering the location of \target{} within 16 days of the burst.}
    \label{tab:obs_list}
\end{table}
\target{} is characterised by a symmetric burst centered on 2023-01-21T01:01:15.4 UTC. Table \ref{tab:obs_list} shows the start and end times of the eight VAST observations covering the position of the burst within 16 days of it occurring, inclusive of the observation it was discovered in. We use the non-detections of repeat bursts in those observations to rule out possible periods.  To do this, we generated an array of possible periods up to 6\,hr, with 10\,s resolution (corresponding to the ASKAP integration time). For each possible period we project whether any repeat bursts would be expected to occur in each observation. We assume a burst width of 110\,s, corresponding to the time above $8\sigma$ significance. If there are any predicted burst times within an observation, or 55\,s before or after, we consider the burst detectable, and hence rule out that period.

Figure \ref{fig:period-constraints} shows the period constraints for each observation for periods up to 2 hours. The bottom line shows the overall constraint after considering the combined constraints of each of the above observations. In short, periods less than 43 minutes are conclusively ruled out and we can also rule out most periods up to $\sim 2\,$h. The constraints above 2\,h are substantially less comprehensive and therefore not reported in the interest of brevity. The more conservative scenario discussed in Section \ref{subsec:classification} combines the constraints from SB47253 and SB47251, i.e. periods up to 18.5 minutes and from 24--37 minutes are ruled out.

\bsp	% typesetting comment
\label{lastpage}
\end{document}